\documentclass[mathematics,article,submit,oneauthor,pdftex]{Definitions/mdpi} 

\firstpage{1} 
\pubvolume{9}
\issuenum{7}
\articlenumber{785}
\pubyear{2021}
\copyrightyear{2021}

\usepackage{amssymb}

\usepackage{graphicx}
\usepackage{subcaption}

\usepackage{enumitem}

\usepackage{listings}
\usepackage{color}

\definecolor{dkgreen}{rgb}{0,0.6,0}
\definecolor{gray}{rgb}{0.5,0.5,0.5}
\definecolor{mauve}{rgb}{0.58,0,0.82}

\Title{Bivariate Infinite Series Solution of Kepler's Equations}

\Author{Daniele Tommasini}

\AuthorNames{Daniele Tommasini}

\address{%
	Applied Physics Department, School of Aeronautic and Space Engineering, Universidade de Vigo, As Lagoas s/n, Ourense, 32004 Spain\\
}

\corres{Correspondence: daniele@uvigo.es}

\begin{document}
\nolinenumbers
{\bf Abstract:}
A class of bivariate infinite series solutions of the elliptic and hyperbolic Kepler equations is described, adding to the handful of 1-D series that have been found throughout the centuries. This result is based on an iterative procedure for the analytical computation of all the higher-order partial derivatives of the eccentric anomaly with respect to the eccentricity $e$ and mean anomaly $M$ in a given base point $(e_c,M_c)$ of the $(e,M)$ plane. Explicit examples of such bivariate infinite series are provided, corresponding to different choices of $(e_c,M_c)$, and their convergence is studied numerically. 
In particular, the polynomials that are obtained by truncating the infinite series up to the fifth degree reach high levels of accuracy in significantly large regions of the parameter space $(e,M)$. Besides their theoretical interest, these series can be used for designing 2-D spline numerical algorithms for efficiently solving Kepler's equations for all values of the eccentricity and mean anomaly.

\vspace{0.2cm}


\vspace{0.2cm}

\noindent
Published in Mathematics 2021, 9(7), 785

\noindent
Published version available at 
\url{https://doi.org/10.3390/math9070785} 
	

\section{Introduction}
\label{intro}

In the Newtonian approximation, 
the time dependence of the relative position of two distant or spherically symmetric bodies that move in each other's gravitational field can be written with explicit analytical formulas involving a finite number of terms only when the eccentricity, $e$, is equal to $0$ or $1$, corresponding to circular and parabolic orbits, respectively \cite{Roy2005}. For $0<e<1$ and for $e>1$, such evolution can be obtained by solving for $E$ one of the following two Kepler Equations (KEs) (see e.g. Chapter 4 of Ref.~\cite{Roy2005}), 
\begin{equation}
	M = f(e, E) = \left\{\begin{matrix} E - e \sin E, \quad \text{for} \quad e<1\cr e\sinh E - E,\quad \text{for} \quad e>1\cr \end{matrix}\right.,
	\label{eq:Keplers}
\end{equation}
where $M$ and $E$ are measures of the epoch and the angular position called the mean and the eccentric anomaly, respectively (for convenience, the same symbols are used here for the elliptic and hyperbolic anomalies, even though they are defined in different ways).

For any given value of $e$, Equations (\ref{eq:Keplers}) can be solved numerically for $E$ by using a root-finding algorithm for the nonlinear equation $f(e,E)-M=0$ (see Ref.~\cite{Colwell1993} for an historical overview). In particular, efficient strategies based on Newton-Raphson iteration method or one its variants have been applied to the elliptic \cite{Danby1983,Conway1986,Odell1986,Charles1998,Palacios2002,Raposo2017,Fukushima1997,Feinstein2006,Zechmeister2018,Zechmeister2021} and hyperbolic \cite{Gooding1988,Fukushima1997b,Avendano2015,Raposo2018} KEs.

Moreover, a handful of infinite series solutions of Eqs. (\ref{eq:Keplers}) have also been found throughout the centuries (see Chapter 3 in Ref.~\cite{Colwell1993}). 
The solution for $0<e<1$ has been written as an expansion in powers of $e$ \cite{Lagrange1771},
or as an expansion in the basis functions $\sin (n M)$ with coefficients proportional to the values $J_n(e)$  of Bessel functions
\cite{Bessel1805,DaSilva1994}.
Levi-Civita \cite{LeviCivita1904a,LeviCivita1904b} described a series in powers of the combination $z=\frac{e\exp(\sqrt{1-e^2})}{1+\sqrt{1-e^2}} $. Finally, Stumpff found an infinite series expansion in powers of $M$ \cite{Stumpff1968}. 

This article describes a class of solutions of KEs, Eqs. (\ref{eq:Keplers}), in terms of bivariate infinite series in powers of both $e$ and $M$,
\begin{equation}
	E =  \sum_{k=0}^\infty \sum_{q=0}^\infty c_{k,q}\, (e-e_c)^k  (M - M_c)^q,
	\label{eq:taylor-bivariate}
\end{equation}
with coefficients $c_{k,q}$ depending on the choice of the base values $(e_c,M_c)$.
These solutions converge locally around $(e_c, M_c)$, and can be used to devise 2-D spline algorithms for the numerical computation of the eccentric anomaly $E$ for every $(e,M)$ \cite{Tommasini2021b}, generalizing the 1-D spline methods that have been described recently  \cite{Tommasini2020a,Tommasini2020b}. Since they do not require the evaluation of transcendental functions in the generation procedure, splines based on polynomial expansions, such as the 1-D cubic spline of Refs. \cite{Tommasini2020a,Tommasini2020b}, or the 2-D quintic spline of Ref.~\cite{Tommasini2021b}, which is based on the solutions presented here, are more convenient for numerical computations than expansions in terms of trigonometric functions.

\section{Methods}
\label{sec:series}

Let the unknown exact solution of Equation (\ref{eq:Keplers}) be
\begin{equation}
	E=g(e,M).
	\label{eq:Inverse_Kepler}
\end{equation}
If the analytical expression of the partial derivatives of $g(e,M)$ were known, 
a bivariate Taylor expansion could be written for any choice of base values $e_c$, $E_c$, $M_c=f(e_c,E_c)$, so that Equation (\ref{eq:taylor-bivariate}) would be demonstrated with the coefficients given by
\begin{equation}
	c_{k,q} = \frac{1}{k! q!} \left[ \frac{\partial^{k+q}  g}{\partial e^k \partial M^q}(e_c, M_c)\right].
	\label{eq:coef-taylor-bivariate}
\end{equation}

To obtain such derivatives, we notice that the definitions in Eqs. (\ref{eq:Keplers}) and (\ref{eq:Inverse_Kepler}) imply the identity,
\begin{equation}
	E = g(e, f(e, E)).
	\label{eq:identity}
\end{equation}
In this expression, $e$ and $E$ are considered as the independent variables. Therefore, by taking the differential, we obtain,
\begin{equation}
	\mathrm{d} E = \frac{\partial g}{\partial e}(e, f(e, E))\, \mathrm{d} e 
	+\frac{\partial g}{\partial M}(e, f(e, E))\left[\frac{\partial f}{\partial e}(e, E) \,\mathrm{d} e + \frac{\partial f}{\partial E}(e, E)\, \mathrm{d} E\right].
	\label{eq:diff-identity}
\end{equation}

Since $e$ and $E$  are independent, the coefficients of $\mathrm{d} E$ and $\mathrm{d} e$ must cancel separately. This condition can be used to obtain the partial derivatives of $g$. Taking also into account Eqs. (\ref{eq:Keplers}) and (\ref{eq:Inverse_Kepler}), the cancellation of the coefficient of $\mathrm{d} E$ implies, 
\begin{equation}
	\frac{\partial g}{\partial M}(e, M) = \frac{1}{\frac{\partial f}{\partial E}\left(e, g(e, M)\right)} \equiv \frac{\lambda}{1-e C}, 
	\label{eq:dgdM}
\end{equation}
where we have defined the parameter $\lambda$ and the functions $C$ such that 
$\lambda=1$ and $C= \cos g(e,M)$, for $e<1$, or $\lambda=-1$ and $C= \cosh g(e,M)$, for $e>1$. 
As it could be expected, Equation (\ref{eq:dgdM}) coincides with the usual rule for the derivative of the inverse function when $e$ is considered to be a fixed parameter \cite{Stumpff1968,Colwell1993}. The cancellation of the coefficient of $\mathrm{d} e$ in Equation (\ref{eq:diff-identity}) implies,
\begin{equation}
	\frac{\partial g}{\partial e}(e, M) =
	-\frac{\partial g}{\partial M}(e, M)\,\frac{\partial f}{\partial e}(e, g(e, M))
	\equiv 
	\frac{S}{1 - e C},
	\label{eq:dgde}
\end{equation}
with $S$ defined as $S= \sin g(e,M)$, for $e<1$, or $S= \sinh g(e,M)$, for $e>1$. In the case of the elliptic KE, the result of Equation (\ref{eq:dgde}) was used in Ref.~\cite{DaSilva1994} to derive an expansion in the basis $\sin n M$.

Eqs.  
(\ref{eq:dgdM})  and (\ref{eq:dgde}), taken together with Eqs. (\ref{eq:Keplers}) and (\ref{eq:Inverse_Kepler}),  can be used for the iterative computation of all the higher order derivatives entering Equation (\ref{eq:taylor-bivariate}) for $e\ne1$. The calculations can be simplified by expressing all the derivatives in terms of only $\lambda$, $S$, and $C$, 
and using the following identities, which can be derived from the definitions of $S$ and $C$ and Eqs. (\ref{eq:dgdM})  and (\ref{eq:dgde}), 
\begin{equation}
	\frac{\partial S}{\partial e}(e, M) = 
	\frac{C\, S}{1 - e\, C},
	\quad    
	\frac{\partial S}{\partial M}(e, M) =  
	\frac{\lambda\, C}{1 - e\, C},
	\label{eq:derivatives-S}
\end{equation}
\begin{equation}
	\frac{\partial C}{\partial e}(e, M) = 
	-\frac{\lambda\, S^2}{1 - e\, C},
	\quad    \frac{\partial C}{\partial M}(e, M) = 
	-\frac{S}{1 - e\, C}.
	\label{eq:derivatives-C}
\end{equation}

The second order derivatives can then be obtained by applying the operators $\frac{\partial}{\partial e}$ and $\frac{\partial}{\partial M}$
and the rules of Eqs.  
(\ref{eq:derivatives-S}) and (\ref{eq:derivatives-C}) to the first order derivatives given in 
Eqs. (\ref{eq:dgdM})  and (\ref{eq:dgde}). The result is,
\begin{equation}
	\frac{\partial^2 g}{\partial e^2} (e, M)
	=    \frac{2\,C\,S}{(1-e\,C) ^2}-\frac{\lambda\,e\, S^3}{(1-e\,C) ^3}, 
	\label{eq:d2gde2}
\end{equation}
\begin{equation}
	\frac{\partial^2 g}{\partial M^2}(e, M)
	=    -\frac{\lambda\,e\,S}{(1-e\,C)^3}, 
	\label{eq:d2gdM2}
\end{equation}
\begin{equation}
	\frac{\partial^2 g}{\partial e \partial M} (e, M)
	=        \frac{\lambda\,C}{(1-e\,C) ^2}-\frac{e\, S^2}{(1-e\,C) ^3}. 
	\label{eq:d2gdedM}
\end{equation}

Similarly, the third order derivatives can be obtained by applying 
the operators $\frac{\partial}{\partial e}$ and $\frac{\partial}{\partial M}$
and the rules of Eqs. 
(\ref{eq:derivatives-S}) and (\ref{eq:derivatives-C}) 
to the second order derivatives, 
Eqs.  (\ref{eq:d2gde2}), (\ref{eq:d2gdM2}), and (\ref{eq:d2gdedM}). 
The result is,
\begin{equation}
	\frac{\partial^3 g}{\partial e^3} (e, M) = \frac{6 \,C^2\, S-3\,\lambda\,S^3}{(1-e\,C) ^3}-\frac{10\, \lambda\, e\, C\, S^3}{(1-e\,C)^4} + \frac{3\, e^2\, S^5}{(1-e\,C)^5}  ,
	\label{eq:dg3de3}
\end{equation}
\begin{equation}
	\frac{\partial^3 g}{\partial M^3}(e, M)
	= 
	-\frac{e\,C}{(1-e\,C)^4} + \frac{3\, \lambda\, e^2\, S^2}{(1-e\,C)^5}  ,
	\label{eq:d3gdM3}
\end{equation}
\begin{equation}
	\frac{\partial^3 g}{\partial e^2 \partial M} (e, M) = 
	\frac{2\,\lambda \,C^2-2\,S^2}{(1-e\,C) ^3}-\frac{7\, e\, C\, S^2}{(1-e\,C)^4} + \frac{3\, \lambda\, e^2\, S^4}{(1-e\,C)^5}  ,
	\label{eq:dg3de2dM}
\end{equation}
\begin{equation}
	\frac{\partial^3 g}{\partial e \partial M^2} (e, M) =
	-\frac{\lambda \,S}{(1-e\,C) ^3}-\frac{4\,\lambda\, e\, C\, S}{(1-e\,C)^4} + \frac{3\,  e^2\, S^3}{(1-e\,C)^5}.
	\label{eq:dg3dedM2}
\end{equation}

All the higher order derivatives can be obtained by iterating this procedure. These expressions are exact, but they depend on the unknown function $g$ through $S$ and $C$. Nevertheless, taken together with Equation (\ref{eq:Inverse_Kepler}), they can be used to compute 2-D Taylor series solutions of KEs. This can be done by choosing a pair of base values, $e_c$ and $E_c$, corresponding to $M_c = f(e_c,E_c)$. The values of the coefficients entering Eqs. (\ref{eq:coef-taylor-bivariate}) and (\ref{eq:taylor-bivariate}) can then be computed 
by substituting
$\lambda=1$, $S=\sin E_c$, $C=\cos E_c$, for $e_c<1$, or $\lambda=-1$, $S=\sinh E_c$, $C=\cosh E_c$, for $e_c>1$, in the expressions for the derivatives of $g$,
and by defining the zeroth order term $ \frac{\partial^0 g}{\partial e^0 M^0}(e_c, M_c) =  g(e_c, M_c) = E_c$. This procedure can be used to build a class of bivariate infinite series solutions of the elliptic and hyperbolic KEs, one for any given choice of base values. Three explicit examples will be given in Section \ref{sec:examples}.

The determination of the radius of convergence for 
the univariate series solutions of KEs has been a formidable mathematical problem (see Chapter 6 of Ref.~\cite{Colwell1993}). In the case of the bivariate series of Equations (\ref{eq:taylor-bivariate}) and (\ref{eq:coef-taylor-bivariate}), the region of convergence in the $(e,M)$ plane can be estimated numerically as discussed in section~\ref{sec:examples}.

\section{Examples, Discussion and Results}
\label{sec:examples}

In this section, three examples of bivariate infinite series solutions of KEs are given. They have been obtained from Eqs. 
(\ref{eq:taylor-bivariate}) and (\ref{eq:coef-taylor-bivariate})
by applying the methods discussed in Section  \ref{sec:series}
for the computation of the derivatives of $g$,  for three different choices of the base values $(e_c,M_c)$. All the non-vanishing terms up to fifth order are shown explicitly. Since $g(e,-M)=-g(e,M)$,  it is sufficient to solve KEs only for positive values of $M$. Moreover,  for $e<1$ the  $M$ domain can be reduced to the interval $0\le M\le \pi$, and then the solution for every $M$ can be obtained by using the periodicity of $f$ and $g$.

In all cases, it is convenient to define approximate solutions $S_n$ obtained by truncating the infinite series of Equation (\ref{eq:taylor-bivariate}) keeping only the terms with $k+q\le n$, so that 
\begin{equation}
	S_n (e,M) =  \sum_{k=0}^n \sum_{q=0}^{n-k} c_{k,q}\, (e-e_c)^k  (M - M_c)^q,
	\label{eq:S_n}
\end{equation}
with coefficients given by Equation  (\ref{eq:coef-taylor-bivariate}).
The errors  $\mathcal{E}_n$ of the approximate solutions $S_n$ can then be evaluated in a self consistent way,
\begin{equation}
	\mathcal{E}_n (e,M) = \vert  S_n(e,M) -  S_n(e,f(e, S_n(e, M))) \vert.
	\label{eq:error_n}
\end{equation}

From a practical point of view, the convergence of the infinite series for certain values of $(e,M)$ means that $\mathcal{E}_n (e,M)$ should tend to decrease for increasing $n$. This idea is used for obtaining an estimate of the region of convergence in the $(e,M)$ parameter space by comparing the average errors for lower and higher values of $n$ with the following condition, 
\begin{equation}
	\mathcal{E}_{1} (e,M) + \mathcal{E}_{2} (e,M) + \mathcal{E}_{3} (e,M) >\frac32 \left[\mathcal{E}_{4} (e,M) + \mathcal{E}_{5} (e,M)\right].
	\label{eq:ruleofthumb} 
\end{equation}

A more refined criterion of convergence can be obtained by studying the scaling behavior of the solutions. 
For this purpose, every point of the $(e,M)$ plane is expressed in terms of polar variables $\rho$, $\phi$, defined as
\begin{equation}
e = e_c + \rho\cos\phi, \qquad M = M_c + \rho\sin\phi. 
\end{equation}
 All the polynomials $S_n(e,M)$ and their errors $\mathcal{E}_{n} (e,M)$ can then be thought of as functions of $\rho$ and $\phi$. For a given value of $\phi$, these functions are one dimensional, depending only on $\rho$, which is a measure of the distance  from the center $(e_c,M_c)$ in the $(e,M)$ plane. Thus $\rho$ can play a role similar  to that of the embedding
parameter $q$ of the homotopy analysis method \cite{Liao2004}, with the difference that $\rho$ will not be assumed to be smaller than $1$. Actually, the parameters $\rho$ and $\phi$ 
will only be used in the intermediate steps and will disappear from the final criteria of convergence. 

\begin{figure}
	\includegraphics[width=\columnwidth]{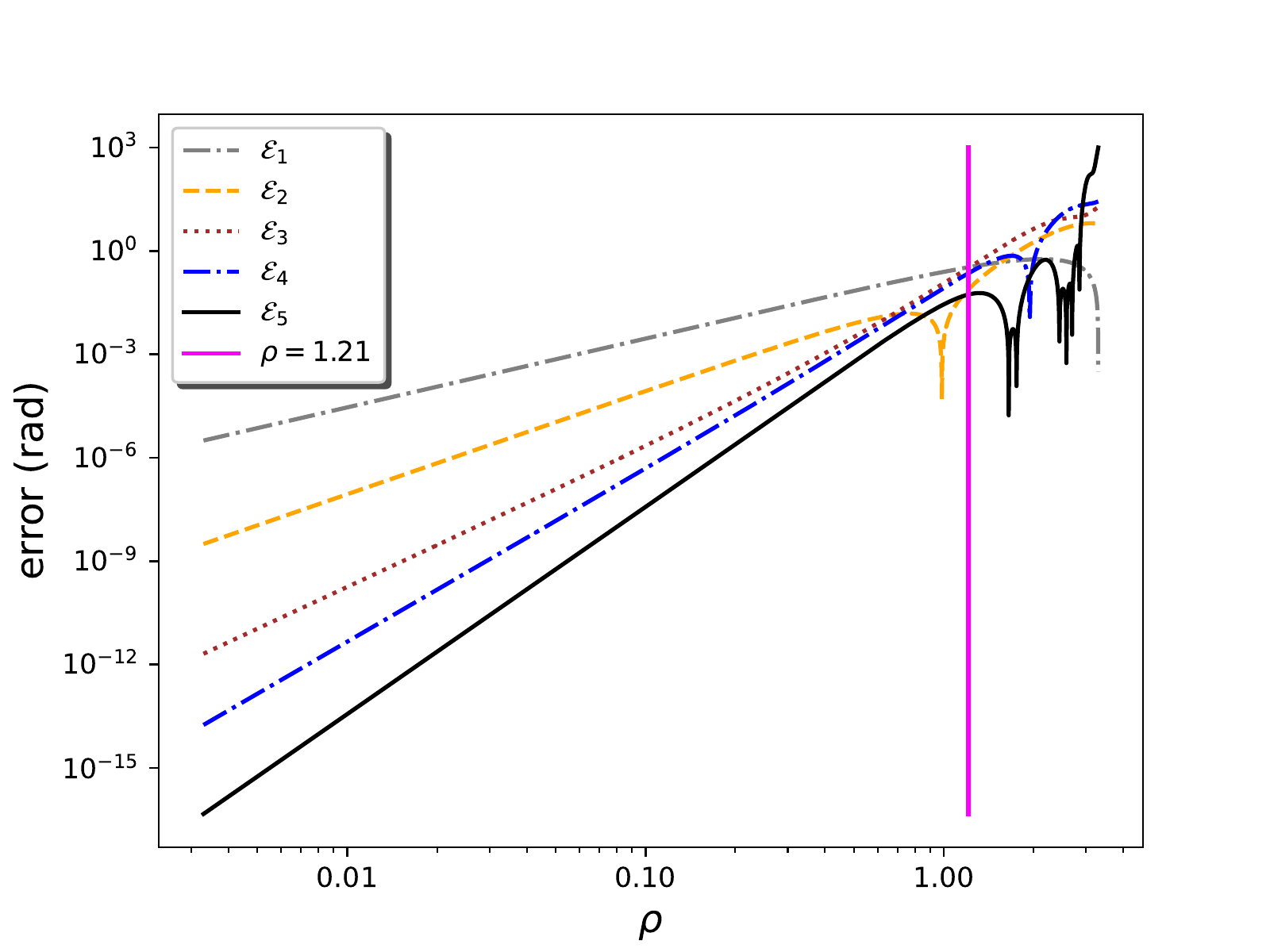}
	\caption{Errors $\mathcal{E}_n(e, M)$ (in logarithmic scale) affecting the approximate polynomial solutions $S_n(e,M)$ of KE for $(e_c,M_c)=(0,0)$
		along the diagonal line $M=\pi e$ of the $(e,M)$ plane (thus $\rho=\sqrt{1+\pi^2}\,e$ and $\tan\phi=\pi$). 
	The $S_n$ are obtained by truncating the infinite series of Equation (\ref{eq:bivariate-series-origin}) up to degree $n$, for $n=1,\cdots,5$. The vertical magenta line at $\rho = 1.21$ corresponds to the  limit below which convergence is obtained in this direction.
	}
	\label{fig:figure1}
\end{figure}

If the bivariate series of Equations (\ref{eq:taylor-bivariate}) and (\ref{eq:coef-taylor-bivariate}) converges in a certain point $(e,M)$ along a fixed direction $\phi$, the error of the $S_n$ approximation can be written as
\begin{equation}
\mathcal{E}_n(e_c+\rho\cos\phi,M_c + \rho\sin\phi) = \frac{\rho^{n+1}}{(n+1)!} \left\vert \frac{\partial^{n+1} g}{\partial \rho^{n+1}}(e_c+\bar\rho\cos\phi,M_c + \bar\rho\sin\phi)\right\vert\equiv \frac{\rho^{n+1}}{(n+1)!}\beta_{n+1}(\phi,\rho),
\label{eq:Taylor-1D-remainder}
\end{equation}
where the derivative entering the definition of $ \beta_{n+1}(\phi,\rho)$ has to be computed for an unknown value $\bar\rho\in[0,\rho]$. 
By plotting the actual numerical errors $\mathcal{E}_n$ in a direction $\phi$
for a given base point $(e_c,M_c)$, it can be seen that the $\rho$ dependence of $\beta_{n+1} $ in the convergence region is usually much milder than that of the factor $\rho^{n+1}$. As an example,  Fig.~\ref{fig:figure1} shows such plots for the series centered around the point $(e_c,M_c)=(0,0)$, choosing the direction identified by the diagonal line $M=\pi e$ (corresponding to $\tan\phi = \pi$).
Along this line, the error $\mathcal{E}_5$ is at the level of arithmetic double precision ($\epsilon_ \text{double} =2.23\times 10^{-16} $) for $e=\frac{M}{\pi}\lesssim 0.0013$. For almost all values of $\rho< 1.21$ (vertical magenta line in Fig.~\ref{fig:figure1}), corresponding to
$e< 0.367$ and $M< 1.15$ rad,
the errors $\mathcal{E}_n$ 
decrease as $n$ increases, as expected for a convergent series, except for the occasional inversion due to cancellations that occur in one of the $S_n$ ($S_2$ around $\rho\sim1$ in the figure). For $\rho> 1.21$, $\mathcal{E}_1$, $\mathcal{E}_3$, and $\mathcal{E}_4$ mix, and the series can be expected to diverge.

Similar results can be obtained for different directions $\phi$ and base points $(e_c,M_c)$. In general, the linear behavior of $\log \mathcal{E}_n$ in the convergence region corresponds  to  $\mathcal{E}_n\propto \rho^{n+1}$ with a very good approximation, so that 
$\mathcal{E}_{n_1}$ and $\mathcal{E}_{n_2}^{(n_1+1)/(n_2+1)}$ scale with the same power of $\rho$. This behavior can be
made more regular by averaging out the possible oscillations that occur in special directions for the individual $\mathcal{E}_n$. This can be done by summing up different $\mathcal{E}_n$ with the corresponding scale exponent, as in the following combinations: 
\begin{equation}
\mathcal{E}_{12}^{\mathrm{sc}} = (\mathcal{E}_{1}^{3/2} + \mathcal{E}_{2})/2, 
\end{equation}
which scales as $\rho^3$, like $\mathcal{E}_{2} $ but with greater regularity; 
\begin{equation}
\mathcal{E}_{123}^{\mathrm{sc}} = (\mathcal{E}_{1}^2 + \mathcal{E}_{2}^{4/3} + \mathcal{E}_{3})/3, 
\end{equation}
which scales as $\rho^4$, like $\mathcal{E}_{3} $;  and
\begin{equation}
\mathcal{E}_{345}^{\mathrm{sc,}} = (\mathcal{E}_{3}^{3/2} + \mathcal{E}_{4}^{6/5} + \mathcal{E}_{5})/3, \qquad \text{and} \qquad  
\mathcal{E}_{45}^{\mathrm{sc}} = (\mathcal{E}_{4}^{6/5} + \mathcal{E}_{5})/2, 
\end{equation}
which scale as $\rho^6$, like $\mathcal{E}_{5} $ but--again--with greater regularity.
Equation (\ref{eq:Taylor-1D-remainder}) and these scaling laws are expected to hold only when the Taylor series converges. Therefore, two additional numerical criteria of convergence are given by the inequalities
\begin{equation}
\mathcal{E}_{345}^{\mathrm{sc}} < \mathcal{E}_{12}^{\mathrm{sc}}, \qquad \text{and} \qquad  
\mathcal{E}_{45}^{\mathrm{sc}} < \mathcal{E}_{123}^{\mathrm{sc}}.
\label{eq:scale-criterion}
\end{equation}
These conditions ensure that the errors not only tend to decrease for increasing $n$, but they also scale as expected when the series is convergent. For the solution based in $(e_c,M_c)=(0,0)$
and evaluated along the diagonal direction $M=\pi e$, these conditions give  the limiting value $\rho=1.21$ shown in Fig.~\ref{fig:figure1}. By inspecting the figure it can be seen that the bounds of Equations (\ref{eq:scale-criterion}) produce a reliable result in this case.
Moreover, as shown in subsection \ref{sec:first-solution}, 
Equations (\ref{eq:scale-criterion}) also reproduce the known radius of convergence of Lagrange series \cite{Lagrange1771,Colwell1993} in the limit where it can be compared with our bivariate series.
The bounds of Equations (\ref{eq:scale-criterion}) are usually more stringent than those obtained from Equation (\ref{eq:ruleofthumb}), but there may be special directions for which the opposite may be true. Hereafter, a conservative definition of the region of convergence will be used by imposing Equations (\ref{eq:scale-criterion}) and Equation (\ref{eq:ruleofthumb}) at the same time.

\subsection{\bf Bivariate Infinite Series Solution of the Elliptic Kepler equation around $e_c=0$, $M_c=0$}
\label{sec:first-solution}

Choosing $e_c=0$, $E_c=0$, so that $M_c=0$ rad, $\lambda_c=1$, $S_c=0$, $C_c=1$, the series of Eqs. (\ref{eq:taylor-bivariate}) and (\ref{eq:coef-taylor-bivariate})  becomes,
\begin{equation}
	E = M + e M + e^2 M + e^3 M - \frac{e}{6} M^3 + e^4 M - \frac{2}{3} e^2 M^3 + \cdots
	\label{eq:bivariate-series-origin}
\end{equation}

This case can be compared with Lagrange's \cite{Lagrange1771} and Stumpff's \cite{Stumpff1968} univariate series, which are,
\begin{multline}
E = M + e \sin M + \frac{e^2}{2} \sin 2 M - \frac{e^3}{8} [\sin M - 3 \sin (3 M)] 
 + \frac{e^4}{6} [-1 + 4 \cos(2 M)] \sin(2 M)+\\
  + \frac{e^5}{192} [23 + 44 \cos(2 M) + 125 \cos(4 M)] \sin(M) + \cdots
 \quad (\mathrm{Lagrange}),
 \label{eq:Lagrange-series}
\end{multline}
and
\begin{equation}
E = \frac{M}{1-e} -  \frac{M^3 e}{3! (1-e)^4} + \frac{M^5 e(9 e +1)}{5! (1-e)^7} + \cdots \quad (\mathrm{Stumpff}),
 \label{eq:Stumpff-series}
 \end{equation}
\citep[see Ref.][Equation (3.25)]{Colwell1993}. It is easy to see that the Taylor expansions (up to fifth order) of Equations (\ref{eq:Lagrange-series}) and (\ref{eq:Stumpff-series}) around $M=0$ and $e=0$, respectively, coincide with the bivariate series of Equation (\ref{eq:bivariate-series-origin}). Of course, their expansions in a neighborhood of $(e_c,M_c)=(0,0)$ have to coincide since all these series solve the same equation around the same point. However, the complete series are different from one another, and their numerical values will also be increasingly different for increasing distance from the base point $(0,0)$. As a consequence, their regions of convergence will also be different.

\begin{figure}
	\includegraphics[width=\columnwidth]{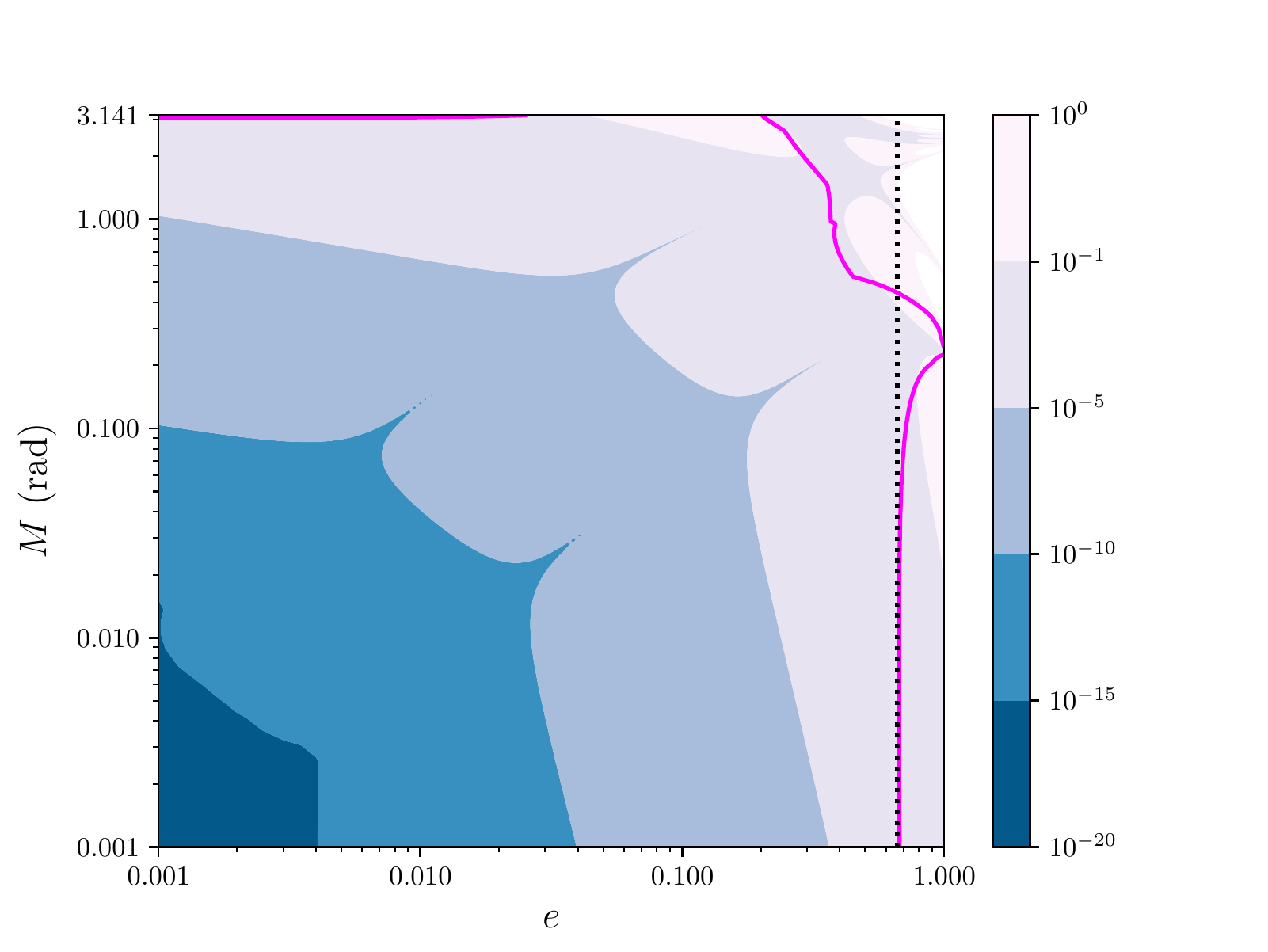}
	\caption{Contour levels of the error $\mathcal{E}_5(e,M)$ affecting the fifth degree polynomial approximation, Equation (\ref{eq:bivariate-series-origin}), as a function of the eccentricity $e$ and the mean anomaly $M$ (both in logarithmic scales). The continuous magenta curve marks the boundary of the region of convergence, as estimated with Equations (\ref{eq:scale-criterion}) and Equation (\ref{eq:ruleofthumb}).
		The vertical dotted line 
		represents the limit 
		of the region of convergence  for Lagrange's 
		univariate series.} 
	\label{fig:figure2}
\end{figure}

Fig. \ref{fig:figure2} shows the contour levels  in the $(e,M)$ plane of the error $\mathcal{E}_5$ affecting the fifth degree polynomial approximation, $S_5$, as given by Equation (\ref{eq:bivariate-series-origin}). The error $\mathcal{E}_5$ is kept below $\sim 10^{-4}$ rad for $e\lesssim 0.5$ and $ M\lesssim \pi/2$, and is reduced to the level  $\sim 10^{-13}$ rad for $e\sim 0.01$ and $M\sim \pi/1000$. Moreover, the fifth order approximation reaches machine precision $\epsilon_\text{double}=2.23\times 10^{-16}$ in an entire neighborhood of size  $\Delta e\sim 2\times 10^{-3}$, $\Delta M\sim3\times 10^{-3}$ rad around the point $(e_c,M_c)$. 
The continuous magenta curve marks the boundary of the region of convergence of the bivariate series of Equation (\ref{eq:bivariate-series-origin}), as estimated with Equations (\ref{eq:scale-criterion}) and Equation (\ref{eq:ruleofthumb}). This can be compared with the limit  $e< 0.6627434193$  for the convergence of Lagrange's univariate series
(see page 26 of Ref.~\cite{Colwell1993}), which is represented by a vertical dotted line in the figure. For $M\ll1$, our limit for the convergence of the bivariate series agrees very well with that of Lagrange's series, as it could be expected since in such regime the first terms of the Taylor expansion for $\sin M$ provide a very good approximation. Not surprisingly, for larger values of $M$ the vertical dotted line separates from the magenta line, so that the region of convergence of the bivariate series is different from that of Lagrange.

\subsection{\bf Bivariate Infinite Series Solution of the Elliptic Kepler Equation around $e_c=\frac12$, $M_c=\frac{\pi-1}{2}$}
\label{sec:second-solution}

Choosing $e_c=\frac12$, $E_c=\frac{\pi}{2}$, so that $M_c=\frac{\pi-1}{2}$, $\lambda_c=1$, $S_c=1$, $C_c=0$, and defining $\delta =e-e_c=e-\frac12$ and $\Delta =M-M_c=M-\frac{\pi-1}{2}$, the series of Eqs. (\ref{eq:taylor-bivariate}) and (\ref{eq:coef-taylor-bivariate})  becomes,	
\begin{multline}
	E = \frac\pi2 + \Delta + \delta -\frac{\delta^2 + 2 \delta\,\Delta + \Delta^2}{4} 
	+\frac{-3\delta^3- 5 \delta^2\Delta - \delta\,\Delta^2+\Delta^3}{8} +\\
	+\frac{85 \delta^4 + 244 \delta^3\Delta + 222 \delta^2\Delta^2+ 52 \delta\, \Delta^3 - 11 \Delta^4}{192} + \\
	+\frac{37 \delta^5 - 35 \delta^4\Delta -318 \delta^3\Delta^2 - 374 \delta^2\Delta^3 -119 \delta\, \Delta^4 + 9 \Delta^5}{384} + \cdots\\
	\label{eq:bivariate-series-middle}
\end{multline}

\begin{figure}
	\includegraphics[width=\columnwidth]{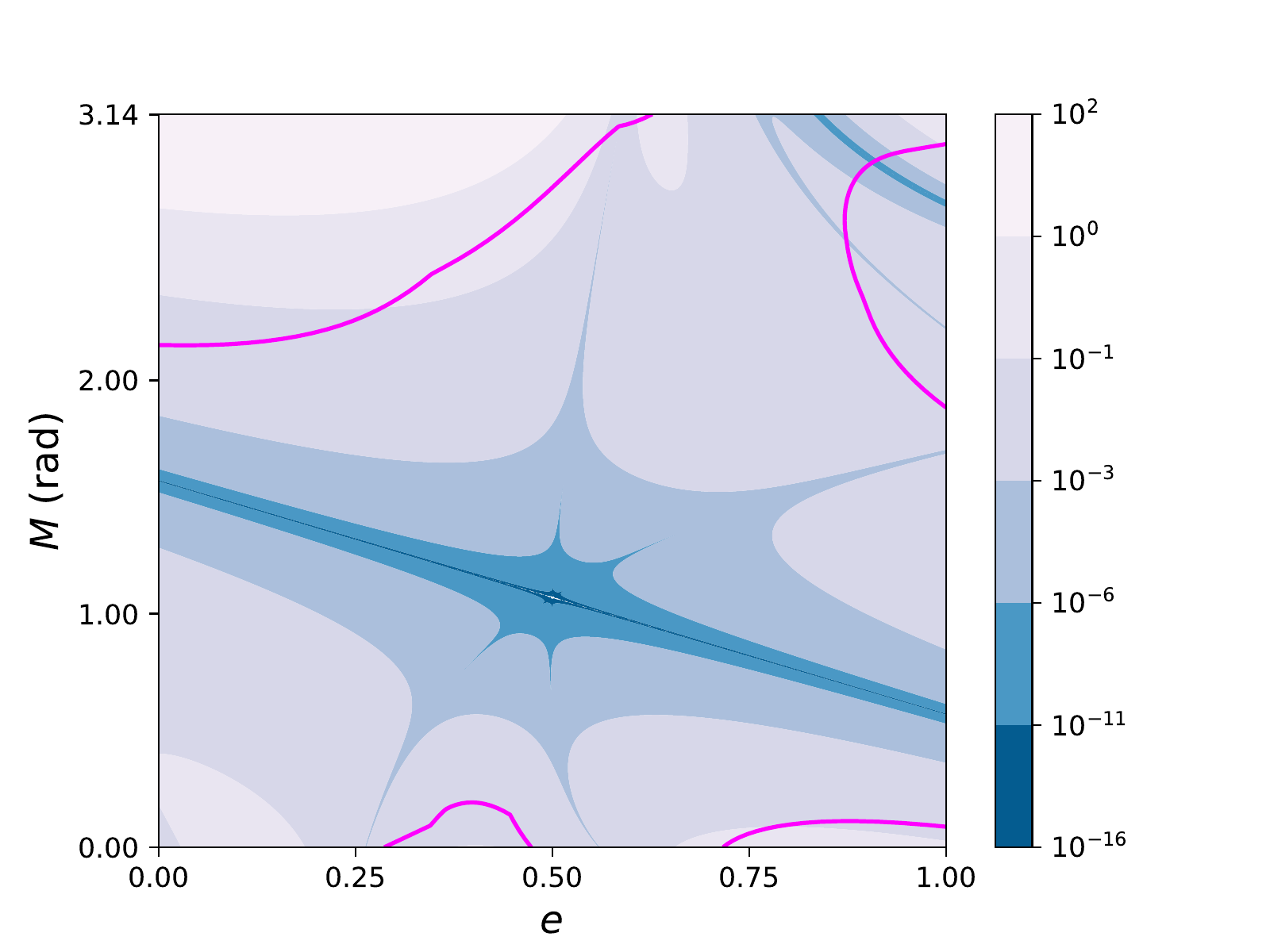}
	\caption{Contour levels of the error $\mathcal{E}_5$ affecting the fifth degree polynomial approximation of Equation (\ref{eq:bivariate-series-middle}), as a function of the eccentricity $e$ and the mean anomaly $M$. The continuous magenta curve marks the boundary of the region of convergence, as estimated with Equations (\ref{eq:scale-criterion}) and Equation (\ref{eq:ruleofthumb}).
(Notice that here the axes for $e$ and $M$ are linear.)}
	\label{fig:figure4}
\end{figure}

Fig. \ref{fig:figure4} shows the contour levels  in the $(e,M)$ plane of the error $\mathcal{E}_5$ affecting the fifth degree polynomial approximation, $S_5$, as given by Equation (\ref{eq:bivariate-series-middle}). 
The continuous magenta curve marks the boundary of the region of convergence, 
as estimated with Equations (\ref{eq:scale-criterion}) and Equation (\ref{eq:ruleofthumb}). It can be seen that the Taylor series based in the mid point $(\frac12,\frac{\pi-1}2)$ converges in a significant part of the $(e,M)$ plane. Moreover, the fifth degree polynomial reaches machine precision $\epsilon_\text{double}$ in an elongated neighborhood of the point $(\frac12,\frac{\pi-1}2)$ along a diagonal line
crossing the entire $e$ domain from $(e=0,M=1.57 \,\text{rad})$ to $(e\simeq 1, M=0.56 \,\text{rad})$, with transverse
size (along the $M$ direction) ranging from  $\sim 3\times 10^{-3}$ rad close to the endpoints, to $\sim10^{-2}$ rad around the center $(e_c,M_c)$.

\subsection{\bf Bivariate Infinite Series Solution of the Hyperbolic Kepler Equation around $e_c=2$, $M_c=0$}
\label{sec:third-solution}

Choosing $e_c=2$, $E_c=0$, so that $M_c=0$, $\lambda_c=-1$, $S_c=0$, $C_c=1$, defining $\delta = e-e_c=e-2$,  the series of Eqs. (\ref{eq:taylor-bivariate}) and (\ref{eq:coef-taylor-bivariate})  becomes,
\begin{equation}
	E =  M -  M \delta + M \delta^2 -\frac{M^3}{3}  - M \delta^3  + \frac{7 M^3\delta}{6}  
	+ M \delta^4 - \frac{8 M^3\delta^2}{3} + \frac{19 M^5}{60}+ \cdots,\\
	\label{eq:bivariate-series-hyperbolic}
\end{equation}
where now $E$ and $M$ indicate the (dimensionless) hyperbolic anomalies.

\begin{figure}
	\includegraphics[width=\columnwidth]{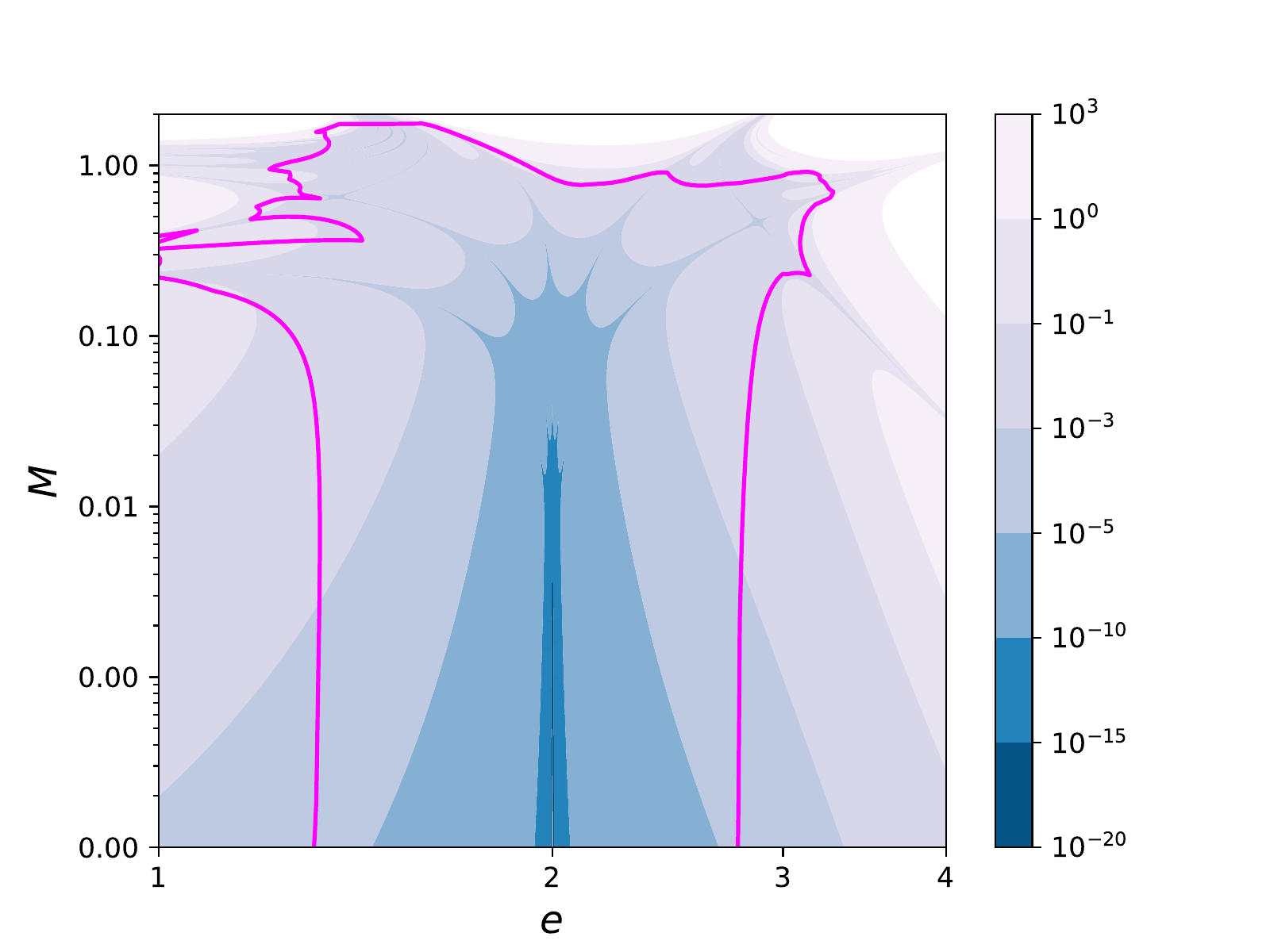}
	\caption{Contour levels of the error $\mathcal{E}_5$ affecting the fifth degree polynomial approximation of Equation (\ref{eq:bivariate-series-hyperbolic}), as a function of the eccentricity $e$ and the mean anomaly $M$ (both in logarithmic scale). 
		The continuous magenta curve marks the boundary of the region of convergence, as estimated with Equations (\ref{eq:scale-criterion}) and Equation (\ref{eq:ruleofthumb}).
	}
	\label{fig:figure6}
\end{figure}

For the hyperbolic motion, the values of $e$ and $M$ can vary in infinite ranges, $1<e<\infty$, $0<M<\infty$ (due to the symmetry for $M\to -M$). In
Fig. \ref{fig:figure6}, the contour levels  in the $(e,M)$ plane of the error $\mathcal{E}_5$ for the solution (\ref{eq:bivariate-series-hyperbolic}) 
have been drawn in the region $e\le 4$, $M\le2$. This region has been chosen in such a way that the plot contains the magenta curve marking the boundary of the region of convergence, 
as estimated with Equations (\ref{eq:scale-criterion}) and Equation (\ref{eq:ruleofthumb}). Moreover, the fifth degree polynomial reaches machine precision $\epsilon_\text{double}$ in an entire neighborhood of size $\Delta e\sim 8\times 10^{-3}$, $\Delta M\sim 2\times 10^{-3}$,  around the point $(e_c,M_c)$.

\section{Conclusions}

I described an analytical procedure for the exact computation of all the higher-order partial derivatives of the elliptic and hyperbolic eccentric anomalies with respect to both the eccentricity $e$ and the mean anomaly $M$. Although such derivatives depend implicitly on the solution of KE, they can be computed explicitly by choosing a couple of base values $e_c$ and $E_c$ for the eccentricity and the eccentric anomaly, so that the corresponding value $M_c$ of the mean anomaly can be obtained without solving KE. For any such choice of  $(e_c,M_c)$, an infinite Taylor series expansion in both $M$ and $e$ can then be written, which is expected to converge in a suitable neighborhood of $(e_c,M_c)$. A 
procedure 
for estimating the actual size of the region of convergence  has also been given.

Three explicit examples of such series were then provided, two for the elliptic and one for the hyperbolic KE. Each of them, for fixed base point, turns out to converge in large parts of the $(e,M)$ plane. 
For $(e,M)$ close to 
$(e_c,M_c)$ within a range $\Delta e\sim\Delta M/\pi = \mathcal{O}(10^{-3})$, the polynomial obtained by truncating the infinite series up to the fifth degree reaches an accuracy at the level of machine double precision. Further away from $(e_c,M_c)$, but still within the region of convergence, higher order terms should be introduced to maintain such an accuracy.

Since these new solutions converge locally around $(e_c, M_c)$, a suitable set of them, centered around different $(e_c, M_c)$ and truncated up to a certain degree, can be used to design an algorithm for the numerical computation of the function $E(e,M)$ for every value of $(e,M)$. The resulting polynomials will form a 2-D spline \cite{Tommasini2021b}, generalizing the 1-D spline that has been proposed in Refs. \cite{Tommasini2020a,Tommasini2020b} 
for solving KE  for every $M$ when $e$ is fixed. This bivariate spline may be used for accelerating computations involving the repetitive solution of Kepler's equation for several different values of $e$ and $M$ \cite{Tommasini2021b}, as for exoplanet search \cite{Makarov2019,Eastman2019}  or for the implementation of Enke's method \cite{Roy2005}.

\section{Acknowledgments}
I thank David N. Olivieri for discussions. This work was supported by grants 67I2-INTERREG, from Axencia Galega de Innovaci\'on, Xunta de Galicia, and FIS2017-83762-P from Ministerio de Economia, Industria y Competitividad, Spain.

\bibliography{bivariate_v2}

\end{document}